\newif\ifpreprint

\documentclass[aps,prl,twocolumn,superscriptaddress]{revtex4}

\usepackage{amsmath,amsfonts}
\usepackage[breaklinks=true,bookmarks=true,hyperfigures=true]{hyperref}

\DeclareMathOperator{\Li}{Li}

\begin{document}

\preprint{Brown-HET-1602}

\title{Classical Polylogarithms for Amplitudes and Wilson Loops}

\author{A.~B.~Goncharov}
\affiliation{Department of Mathematics, Brown University,
          Box 1917, Providence, Rhode Island 02912, USA}
\author{M.~Spradlin}
\author{C.~Vergu}
\author{A.~Volovich}
\affiliation{Department of Physics, Brown University,
          Box 1843, Providence, Rhode Island 02912, USA}

\begin{abstract}
We present a compact analytic formula
for the two-loop six-particle
maximally helicity violating remainder function (equivalently, the two-loop lightlike
hexagon Wilson loop) in ${\cal N} = 4$ supersymmetric Yang-Mills theory
in terms of the classical polylogarithm functions $\Li_k$
with cross-ratios of momentum twistor invariants as their arguments.
In deriving our formula we rely on results from the theory of motives.
\end{abstract}

\maketitle

\section{Introduction}

The past few years have witnessed revolutionary advances in our
understanding of the structure of scattering amplitudes, especially
in ${\cal N} = 4$ supersymmetric Yang-Mills theory (SYM).  It is
easy to argue that the seeds of modern progress were sown already
in the 1980s with the discovery of the Parke-Taylor formula
for the simplest
nontrivial amplitudes:
tree-level maximally helicity violating (MHV) gluon scattering.
The mere existence of such a simple formula for
a quantity which otherwise would have been prohibitively difficult to
calculate using
traditional Feynman diagram methods
signalled the tantalizing possibility that a great vista of
unanticipated structure in scattering amplitudes awaited exploration.

In contrast to the situation at tree level,
it is fair to say that recent
progress at loop
level has mostly been evolutionary rather than revolutionary, driven
primarily by faster computers, improved algorithms (both analytic and
numeric), and
software for multiloop calculations
which has been made publicly available.
Yet we hope
that a great new vista of unexplored structure awaits us also
at loop level in
SYM theory.

This paper is concerned with the planar two-loop six-particle
MHV amplitude~\cite{Bern:2008ap,Cachazo:2008hp},
which in a sense is the simplest
nontrivial SYM loop amplitude.
The known infrared and collinear behavior of general amplitudes, conveniently
encapsulated in the ABDK/BDS ansatz~\cite{Anastasiou:2003kj,Bern:2005iz},
determines the $n$-particle
MHV amplitude at each loop order $L \ge 2$
up to an additive finite function of kinematic invariants
called the remainder
function $R_n^{(L)}$.  Given the presumption of
dual conformal invariance~\cite{Drummond:2007au,Drummond:2008vq} for
SYM amplitudes (not yet proven, but supported
by all available evidence~\cite{Anastasiou:2003kj,Bern:2005iz,Cachazo:2006tj,Bern:2006vw,Bern:2008ap}), $R_n^{(L)}$ can depend on conformal
cross-ratios only.  Since there are no cross-ratios for $n=4,5$, the
first nontrivial
remainder function is $R_6^{(2)}$.

The same function $R_6^{(2)}$ is also
believed~\cite{Alday:2007hr,Drummond:2007aua,Brandhuber:2007yx,Berkovits:2008ic}
to arise
as the expectation value of the two-loop lightlike hexagon Wilson loop
in SYM theory~\cite{Drummond:2007bm,Drummond:2008aq}
(after appropriate subtraction of ultraviolet divergences,
e.g.~\cite{Anastasiou:2009kna}).
Numerical agreement between the two remainder functions
was established in~\cite{Bern:2008ap,Drummond:2008aq}.  In a heroic effort,
Del Duca, Duhr, and Smirnov
(DDS)
explicitly evaluated the appropriate Wilson loop diagrams to obtain an
analytic expression
for $R_6^{(2)}$ as a 17-page linear combination of generalized
polylogarithm functions~\cite{DelDuca:2009au,DelDuca:2010zg}
(see also~\cite{Zhang:2010tr}).

The motivation for the present work is the belief that
if SYM theory is really
as beautiful and rich as recent developments indicate, then there
must exist a more enlightening
way of expressing the remainder function $R_6^{(2)}$.
Ideally, like the Parke-Taylor formula at tree level,
the expression should provide encouragement and guidance
as we seek deeper understanding of SYM at loop level.

We present our new formula for $R_6^{(2)}$ in the next section and
then describe the algorithm by which it was obtained.

\section{\label{sec:remainder-function}The Remainder Function $R_6^{(2)}$}

The remainder function $R_6^{(2)}$ is usually presented as a
function of
the three dual conformal cross-ratios
\begin{equation}
\label{eq:crossratios}
u_1 = \frac{s_{12} s_{45}}{s_{123} s_{345}}, \quad
u_2 = \frac{s_{23} s_{56}}{s_{234} s_{123}}, \quad
u_3 = \frac{s_{34} s_{61}}{s_{345} s_{234}},
\end{equation}
of the momentum invariants $s_{i\cdots j} = (k_i + \cdots + k_j)^2$,
though we will see shortly that cross-ratios of momentum twistor invariants
are more natural variables.
In terms of
\begin{equation}
x^\pm_i = u_i x^\pm, \qquad
x^\pm = \frac{u_1+u_2+u_3-1 \pm \sqrt{\Delta}}{2 u_1 u_2 u_3},
\end{equation}
where $\Delta = (u_1+u_2+u_3-1)^2 - 4 u_1u_2u_3$,
we find
\begin{multline}
\label{eq:mainresult}
R^{(2)}_6(u_1,u_2,u_3) = \sum_{i=1}^3 \left( L_4(x^+_i, x^-_i) -
\frac{1}{2} \Li_4(1 - 1/u_i)\right) \cr
- \frac{1}{8} \left( \sum_{i=1}^3 \Li_2(1 - 1/u_i) \right)^2
+ \frac{1}{24}J^4 +  \frac{\pi^2}{12} J^2 + \frac{\pi^4}{72}.
\end{multline}
Here we use the functions
\begin{multline}
\label{eq:bwrz}
L_4(x^+, x^-) =
\frac{1}{8!!} \log(x^+ x^-)^4
\cr
+
\sum_{m=0}^3
\frac{(-1)^m}{(2m)!!} \log(x^+ x^-)^m
(\ell_{4-m}(x^+) + \ell_{4-m}(x^-))
\end{multline}
and
\begin{equation}
\ell_n(x) = \frac{1}{2} \left( \Li_n(x) - (-1)^n \Li_n(1/x) \right),
\end{equation}
as well as the quantity
\begin{equation}
J = \sum_{i=1}^3 (\ell_1(x^+_i) - \ell_1(x^-_i)).
\end{equation}
Note that in the Euclidean region
where all $u_i > 0$, the $x_i^+$ never enter the lower half-plane
and the $x_i^-$ never enter the upper half-plane.
The expression~(\ref{eq:mainresult}) is valid
in the Euclidean region
with the understanding that the branch
cuts of $\Li_n(x_i^+)$ and $\Li_n(1/x_i^-)$ are taken to lie below
the real axis while the branch cuts of $\Li_n(x_i^-)$ and $\Li_n(1/x_i^+)$
are taken to lie above the real axis.
(The quantities $x_i^+ x_i^-$ appearing as arguments of the $\log$s
are always positive.)
In writing~(\ref{eq:mainresult})
extreme care has necessarily been taken to ensure the proper
analytic structure.  For example one can easily check that
$J$ naively simplifies to $\frac{1}{2} \log(x^-/x^+)$, but this relation
only holds in the regions $\Delta > 0$ or $u_1+u_2+u_3<1$.
We caution the reader that any attempt to use any such naive relations,
including the well-known relation between $\Li_n(1/x)$ and
$\Li_n(x)$, without careful consideration of the branch structure,
voids our warranty on~(\ref{eq:mainresult}).

Besides its great simplicity,
two
notable features of~(\ref{eq:mainresult}) which set it apart
from the DDS formula are manifest symmetry under any
permutation of the $u_i$, and the fact that the expression is
valid and
readily evaluated for all positive $u_i$, in particular also
outside the
unit cube.

\section{\label{sec:description}Description of the Algorithm}

\subsection{A Convenient Choice of Variables}

The DDS formula is expressed in terms of the classical
polylogarithms $\Li_k$ as well as
a collection of considerably more complicated multiparameter generalizations
studied by one of the authors~\cite{Goncharov} and
defined recursively by
\begin{equation}
G(a_k,a_{k-1},\ldots;z) = \int_0^z G(a_{k-1},\ldots;t) \frac {d t}{t - a_k}
\end{equation}
with $G(z) \equiv 1$, of which the harmonic polylogarithms
familiar in the physics literature~\cite{Remiddi:1999ew} are special cases.

The parameters of the various transcendental functions which appear
in the DDS formula involve
not just the cross-ratios~(\ref{eq:crossratios}), but also the
more complicated
combinations $1 - u_i$, $(1 - u_i)/(1-u_i-u_j)$, $u_i+u_j$,
$
u_{jkl}^\pm = \frac{1 - u_j - u_k + u_l \pm \sqrt{\Delta}}{2 (1-u_j) u_l},
$
and
$v_{jkl}^\pm = \frac{u_k - u_l \pm
\sqrt{(u_k + u_l)^2 - 4 u_j u_k u_l}}{2(1-u_j) u_k}$.
This large collection of variables is redundant in an
inefficient way, with many rather complicated algebraic identities
amongst them.

Our computation is greatly facilitated by
a judicious choice of variables which trivializes all of these
algebraic relations.  We choose to express the three $u_i$ by six variables
$z_i$ valued in $\mathbb{P}^1$ (with an $SL(2,{\mathbb{C}})$ redundancy) via
\begin{equation}
\label{eq:diagonals}
u_1 = \frac{z_{23} z_{56}}{z_{25} z_{36}}, \quad
u_2 = \frac{z_{16} z_{34}}{z_{14} z_{36}}, \quad
u_3 = \frac{z_{12} z_{45}}{z_{14} z_{25}},
\end{equation}
where $z_{ij} = z_i - z_j$.  One virtue of these coordinates
is that $\Delta$ becomes a perfect
square, so that the $u^\pm_{jkl}$ are rational functions
of the $z_{ij}$.  (The $v^\pm_{jkl}$ completely drop out
as explained
in the following subsection.)

We anticipate that for general $n$ the best variables for studying
the remainder function will be the momentum twistors of~\cite{Hodges:2009hk}.
Indeed the $z$ variables may be thought of as a particular simplification
of momentum twistors which is valid for the special case $n=6$ via
the relation
$\langle a b c d \rangle \propto z_{ab} z_{ac} z_{ad} z_{bc} z_{bd} z_{cd}$.
In terms of momentum twistors
\begin{equation}
u_1 = \frac{\langle 1234 \rangle \langle 4561 \rangle}
{\langle 1245 \rangle \langle 3461 \rangle}, ~~~~
x_1^+ = - \frac{\langle 1456 \rangle \langle 2356 \rangle}
{\langle 1256 \rangle \langle 3456 \rangle}, \quad {\rm etc}.
\end{equation}

\subsection{The Symbol of a Transcendental Function}

We define a function $T_k$ of transcendentality degree $k$ as one which
can be written as a linear combination (with rational coefficients)
of $k$-fold iterated integrals of the form
\begin{equation}
  \label{eq:iterated}
  T_k = \int_a^b d \log R_1 \circ \cdots \circ d \log R_k,
\end{equation}
where $a$ and $b$ are rational numbers, $R_i(t)$ are
rational functions with rational coefficients and the iterated
integrals are defined recursively by
\begin{multline}
  \int_a^b d \log R_1 \circ \cdots \circ d \log R_n =\\ \int_a^b
  \left(\int_a^t d \log R_1 \circ \cdots \circ d \log R_{n-1}\right) d
  \log R_n(t).
\end{multline}
The integrals are taken along paths from $a$ to $b$.
When the $R_i$ are rational functions
in several variables
the issue of local path independence (or homotopy invariance)
is important (see~\cite{Goncharov:2009}),
and we have checked that $R_6^{(2)}$ has this property.

A useful quantity associated with $T_k$ is its symbol, an element
of the $k$-fold tensor product of the multiplicative group of
rational functions modulo constants
(see~\cite[sec.~3]{Goncharov:2009}).  The symbol of the function
shown in~(\ref{eq:iterated}) is
\begin{equation}
  \label{eq:symbol-definition}
  \text{symbol}(T_k)= R_1 \otimes \cdots \otimes R_k,
\end{equation}
and this definition is extended to all functions of degree $k$ by linearity.

The group property for rational functions $R_i$ modulo constants
implies that
\begin{gather}
  \begin{split}
R_1 \cdots \otimes (R_a R_b) \otimes \cdots R_k
= R_1 \cdots \otimes R_a \otimes \cdots R_k
\\
+ R_1 \cdots \otimes R_b \otimes \cdots R_k,
  \end{split}\\
R_1 \cdots \otimes (c R_a) \otimes \cdots R_k
= R_1 \cdots \otimes R_a \otimes \cdots R_k,
\end{gather}
for any constant $c$ and rational functions $R_i$.

The symbol of all functions appearing in the DDS expression for
the remainder function $R_6^{(2)}$ are readily computed using the above
definitions.
For example the classical polylogarithm functions
\begin{equation}
\Li_k(z) = \int_0^z \Li_{k-1}(t) d\log t, \quad
\Li_1(z) = - \log(1-z)
\end{equation}
clearly have
\begin{equation}
\label{eq:lisymbol}
{\rm symbol}(\Li_k(z)) = - (1 - z) \otimes
\underbrace{z \otimes \cdots \otimes z}_{k-1~{\rm times}}.
\end{equation}

Finally, we note that knowledge of the symbol does not uniquely
determine a transcendental function; there is an ambiguity consisting
in adding lower degree functions multiplied by numerical
constants of the appropriate degree.

\subsection{Constructing a Prototype}

Our first goal is to construct a simple function which
has the same symbol $S$ as that of the DDS formula.
Interestingly we find that the $v_{jkl}^\pm$ variables
completely drop out of the symbol, even before
switching to the $z$ variables.
Although the full expression is too lengthy to reproduce
here (see~\cite{EPAPS}),
we find that $S$
possesses a crucial and surprising symmetry property which in components (the
subscripts here run over the 15 different $z_{ij}$'s) reads
\begin{equation}
\label{eq:symmetry}
\left[ S_{abcd} - S_{bacd} - S_{abdc} + S_{badc} \right]
- (a \leftrightarrow c, b \leftrightarrow d) = 0.
\end{equation}

It is known (see~\cite{Goncharov2}) that transcendental functions of degree
less than four can always be expressed in terms of logarithms and the
classical polylogarithm functions $\Li_k$ only.
However an arbitrary function of degree four
(or higher) does not have this property,
so it is rather remarkable that the remainder
function $R_2^{(6)}$ can be so
expressed.

The explanation for this miracle is Conjecture~1.19
of~\cite{Goncharov2}, from which it follows that
any symbol obeying~(\ref{eq:symmetry})
can be obtained from a function involving
only logarithms and the classical polylogarithm
functions $\Li_k$ with $k \leq 4$.
There is however no {\it ab initio} statement one
can make about the arguments of those polylogarithms; they could
in principle be arbitrary algebraic functions of the
variables appearing in the symbol.

The key to constructing the function from the knowledge of its symbol
lies in identifying the symmetry properties of the five kinds of
terms which can appear, shown in
the following table:

\begin{center}
\begin{tabular}{c|c|c|c|c|}
Function & A$\otimes$A & S$\otimes$A & A$\otimes$S & S$\otimes$S \\ \hline
$\Li_4(x)$ & - & - & \checkmark & \checkmark \\
$\Li_3(x) \log(y)$ & - & - & \checkmark & \checkmark \\
$\Li_2(x) \Li_2(y)$ & \checkmark & \checkmark & \checkmark & \checkmark \\
$\Li_2(x) \log(y) \log(z) $ & - & \checkmark & \checkmark & \checkmark \\
$\log(x)\log(y)\log(z)\log(w)$ & - & - & - & \checkmark \\
\end{tabular}
\end{center}
\noindent
Here A (S) stands for antisymmetric (symmetric), and
a checkmark under A$\otimes$A indicates that the function's symbol contains a piece
which is antisymmetric under exchange of the first two arguments and antisymmetric
under exchange of the last two arguments, etc.  The function $\log(x)
\log(y) \log(z) \log(w)$
has a fully
symmetric symbol and hence only contributes to S$\otimes$S.

The above table suggests the procedure for constructing a function
with the desired symbol $S$.  First we isolate the A$\otimes$A part of $S$
and write down a linear combination of terms of the form $\Li_2(x) \Li_2(y)$
with that symbol, which is only possible if~(\ref{eq:symmetry})
is satisfied.  After subtracting
the symbol of this function from $S$
we isolate the $S \otimes A$ part which we fit with functions
of the form $\Li_2(x) \log(y) \log(z)$.  At the next step, A$\otimes$S, we need
both $\Li_4(x)$ and $\Li_3(x) \log(y)$, which can be
isolated from each other because the former does not contribute to the
part of the symbol which is antisymmetric under exchange of the middle
two arguments, as evident in~(\ref{eq:lisymbol}).
Ultimately we are left with terms which are completely symmetric under
exchange of all arguments and can be fit via functions of the form
$\log(x) \log(y) \log(z) \log(w)$.

This procedure leads to
a function
of degree 4 whose symbol matches that of
the DDS function
using only classical polylogarithm functions.  One interesting
feature which emerges
is that the arguments of all polylogarithms in the result are either
``diagonal'' cross-ratios such as~(\ref{eq:diagonals}), which omit
two diagonally opposite $z$'s,
or ``edge'' cross-ratios, which omit two neighboring $z$'s, such as
\begin{alignat}{3}
\label{eq:edges}
x^-_1&= - \frac{z_{14} z_{23}}{z_{12} z_{34}}, ~~
x^-_2&= - \frac{z_{16} z_{25}}{z_{12} z_{56}}, ~~
x^-_3&= - \frac{z_{36} z_{45}}{z_{34} z_{56}}, \\
x^+_1&= - \frac{z_{14} z_{56}}{z_{16} z_{45}}, ~~
x^+_2&= - \frac{z_{25} z_{34}}{z_{23} z_{45}}, ~~
x^+_3&= - \frac{z_{12} z_{36}}{z_{16} z_{23}}.
\end{alignat}

\subsection{Fitting the Remainder Function $R_6^{(2)}$}

The prototype for $R_6^{(2)}$ is incomplete in two related aspects.
First of all there is considerable ambiguity in the placement of branch
cuts, which the symbol (since it involves only rational
functions) is completely insensitive to.
For example one may be tempted to use the identity
\begin{equation}
\label{eq:bad}
\prod_{i=1}^3 \frac{1 - x_i^+}{1 - x_i^-} = (x^+/x^-)^2
\end{equation}
to replace $J$ by $\frac{1}{2} \log(x^-/x^+)$
as mentioned above,
but the two quantities have
different branch structures in $u$-space.
All such ambiguities are resolved by the physical requirement
that
the remainder function should be
real-valued and smooth
in the Euclidean region.

Secondly it remains to determine the terms
in~(\ref{eq:mainresult}) proportional to $\pi^2$ or $\pi^4$,
which do not
contribute to the symbol.  These terms can be fixed by several considerations
such as checking
the various Regge limits which were tabulated in~\cite{DelDuca:2010zg} (see
also~\cite{Brower:2008nm,Bartels:2008ce,Brower:2008ia,DelDuca:2008jg,Schabinger:2009bb}),
the
requirement that $R_6^{(2)}$ must vanish in the collinear limit
$R_6^{(2)}(1-u,u,0) = 0$, or
by numerically fitting to the DDS formula.  We
have also checked agreement with the leading term in the expansion around the collinear limit computed in~\cite{Alday:2010ku}.

\section{\label{sec:discussion}Discussion}

We have presented in~(\ref{eq:mainresult}) a compact analytic expression
for the two-loop six-particle MHV
amplitude or Wilson loop remainder function
$R_6^{(2)}$ in terms of the classical polylogarithm functions $\Li_k$ only.
The simplicity of our result suggests that it is no
longer outrageous to
imagine the possibility
of determining the remainder function for all
$n$ and any number $L$ of loops, with an aim of matching onto
the work
of~\cite{Alday:2009ga,Alday:2009yn,Alday:2009dv,Alday:2010vh,Alday:2010ku}
at strong coupling.

For $n>6$ we imagine that the remainder function
will continue to involve only the diagonal and edge momentum twistor
cross-ratios, generalizing~(\ref{eq:diagonals}) and~(\ref{eq:edges}).
It is also notable that our formula involves a particular
combination~(\ref{eq:bwrz})
of polylogarithm functions with special analytic properties which is
evidently very closely
related to the Bloch-Wigner-Ramakrishnan-Zagier (BWRZ) functions
(see for example~\cite{Zagier}).  These generalize very
easily to any even degree $2L$, suggesting
a natural appearance in the $L$-loop remainder function.
We anticipate that
the consistency of
collinear limits, and especially the systematic
expansion away from them recently studied in~\cite{Alday:2010ku},
will tightly constrain the general form of $R_n^{(L)}$.

Finally we cannot pass up the opportunity to point out that
the Bloch-Wigner functions (the degree 2 case of the BWRZ functions)
are known~\cite{Zagier} to compute the volumes of hyperbolic
tetrahedra, raising the fascinating possibility of making
a connection to recent work relating
scattering amplitudes to volumes of polytopes in
AdS${}_5$~\cite{Gorsky:2009nv,Mason:2010pg}
(see also~\cite{Hodges:2010kq,Arkani-Hamed:2010}).

\section*{Acknowledgments}

We have benefitted from stimulating discussions with N.~Arkani-Hamed,
J.~Maldacena, and E.~Witten.  CV is grateful to Humboldt University for
hospitality during the course of this work,
which was supported in part by the
DOE under contract DE-FG02-91ER40688 Tasks A (AV) and J OJI (MS),
and by the NSF under grants
PECASE PHY-0643150 and
ADVANCE 0548311 (AV).

\end{document}